\documentclass[twocolumn,pre,superscriptaddress,amsmath,amssymb,longbibliography]{revtex4-1}
\usepackage{hyperref}
\usepackage{graphicx}% Include figure files
\usepackage{dcolumn}% Align table columns on decimal point
\usepackage{bm}% bold math
\usepackage{float}
\usepackage{color}
\begin{document}

%\preprint{Applied Physics Review}

\title{Structure-Property Relationship in Disordered Hyperuniform Materials: Microstructure Representation, Field Fluctuations and Effective Properties}

\author{Liyu Zhong}
\affiliation{Materials Science and Engineering, Arizona State
University, Tempe, AZ 85287}
\affiliation{Department of Mechanics and Engineering Science, College of Engineering, Peking University, Beijing 100871, P. R. China}
\author{Sheng Mao}
\email[correspondence sent to: ]{maosheng@pku.edu.cn}
\affiliation{Department of Mechanics and Engineering Science, College of Engineering, Peking University, Beijing 100871, P. R. China}
%\author{David Keeney}
%\affiliation{Materials Science and Engineering, Arizona State University, Tempe, AZ 85287}
%\author{Duyu Chen}
%\email[correspondence sent to: ]{duyu@alumni.princeton.edu}
%\affiliation{Materials Research Laboratory, University of California, Santa Barbara, California 93106, United States}
\author{Yang Jiao}
\email[correspondence sent to: ]{yang.jiao.2@asu.edu}
\affiliation{Materials Science and Engineering, Arizona State
University, Tempe, AZ 85287} \affiliation{Department of Physics,
Arizona State University, Tempe, AZ 85287}

%shall we also add Pinshine and Vojetch?


\date{\today}% It is always \today, today,

\begin{abstract}
Disordered hyperuniform (DHU) materials are an emerging class of exotic heterogeneous material systems characterized by a unique combination of disordered local structures and a hidden long-range order, which endow them with unusual physical properties, including large isotropic photonic band gaps, superior resistance to fracture, and nearly optimal electrical and thermal transport properties, to name but a few. Here, we consider material systems possessing continuously varying local material properties $\mathcal{K}({\bf x})$ (e.g., thermal or electrical conductivity), modeled via a random field. We devise quantitative microstructure representation of the material systems based on a class of analytical spectral density function ${\tilde \chi}_{_\mathcal{K}}({k})$ associated with $\mathcal{K}({\bf x})$, possessing a power-law small-$k$ scaling behavior ${\tilde \chi}_{_\mathcal{K}}({k}) \sim k^\alpha$. By controlling the exponent $\alpha$ and using a highly efficient forward generative model, we obtain realizations of a wide spectrum of distinct material microstructures spanning from hyperuniform ($\alpha>0$) to nonhyperuniform ($\alpha=0$) to antihyperuniform ($\alpha<0$) systems. Moreover, we perform a comprehensive perturbation analysis to quantitatively connect the fluctuations of the local material property to the fluctuations of the resulting physical fields. In the weak-contrast limit, i.e., when the fluctuations of the property are much smaller than the average value, our first-order perturbation theory reveals that the physical fields associated with Class-I hyperuniform materials (characterized by $\alpha \ge 2$) are also hyperuniform, albeit with a lower hyperuniformity exponent ($\alpha-2$). As one moves away from this weak-contrast limit, the fluctuations of the physical field develop a diverging spectral density at the origin, revealed by our higher-order analysis. We also establish an end-to-end mapping connecting the spectral density of the local material property to the overall effective conductivity of the material system via numerical homogenization. We observe a sharp decrease of the variance of effective properties across realizations as $\alpha$ increases from antihyperuniform values to hyperuniform values. Our results have significant implications for the design of novel DHU materials with targeted physical properties.

%Although not constrained in the reconstructions, we find the composite micrsotuctures with, phase volume fraction $\phi \in [0.4, 0.6]$ and the size of Omega considered here, possess the property of phase-inversion symmetry, which leads to percolation threshold $\phi_c = 0.5$.  

\end{abstract}

%\keywords{Disorder, Hyperuniform, Nanotubes, 2D Materials, High/Medium-Entropy Alloys}

%Use showkeys class option if keyword
%Hyperuniform                            %display desired

\maketitle

%\tableofcontents

%\newpage
%\newpage
\section{Introduction}

A wide class of engineering materials such as composites, alloys, porous media, granular matters, which are of great importance in a diverse spectrum of applications from soft gripping \cite{nguyen2023liquid, elango2015review, mohammadi2021robust} to wave manipulations \cite{waveguide, damaskos1982dispersion, itin2010dispersion, kim2023extraordinary}, typically possess disordered heterogeneous microstructures \cite{torquato2002random, Sa03a}. The complex microstructure space poses many challenges for the design and optimization of heterogeneous material systems via classic approaches such as topology optimization \cite{sigmund1999design, sigmund1997design}. Recently, alternative materials informatics approaches \cite{courtright2025high, generale2024inverse,liu2024active, bessa2017framework,li2022machine, mirzaee2025inverse} have been developed, a key of which is the construction of a set of concise microstructure representations in the reduced-dimension latent space. The original microstructure space is then encoded into the latent space, based on which analytical \cite{cang2018improving, cheng2022data} or data-driven \cite{xu2022correlation} structure-property relationships are established for material design. Subsequently, the optimized representations are decoded to obtain explicit realizations of material microstructure achieving the targeted properties, which is also referred to as the microstructure construction process \cite{bostanabad2018computational, sahimi2021reconstruction,Ye98a, Ye98b}.

Among existing microstructure representations \cite{roberts1997statistical, niezgoda2008delineation, okabe2005pore, jiao2007modeling, jiao2008modeling, hajizadeh2011multiple, tahmasebi2013cross, tahmasebi2012multiple, xu2013stochastic, xu2014descriptor, cang2017microstructure, yang2018microstructural,li2018transfer, farooq2018spectral, cheng2022data, xu2022correlation, skolnick2024quantifying, shih2024fast, casiulis2024gyromorphs}, the spatial correlation functions (SCFs) \cite{To02a, jeulin2021morphological}, especially the lower-order correlation functions \cite{jiao2007modeling, jiao2008modeling, jiao2009superior, chen2019hierarchical, chen2020probing} and the spectral density functions \cite{farooq2018spectral, Ch18a, shi2023computational, shi2025three} have been widely employed to model a variety of heterogeneous material systems \cite{jiao2010geometrical_a, gerke2015improving, karsanina2018hierarchical, feng2018accelerating, jiao2013modeling, chen2015dynamic, jiao2014modeling, guo2014accurate, chen2016stochastic, gerke2019calculation, karsanina2022stochastic, chen2022quantifying, postnicov20232, postnicov2024evaluation}, due to their superior explainability \cite{cheng2022data} and rigorous connection to the physical properties of the materials via analytical contrast expansion formalisms \cite{torquato1997effective, torquato1985effective, kim2020effective, torquato2021nonlocal, kim2023effective, torquato2021diffusion, skolnick2023simulated, torquato1990rigorous, torquato2020predicting, skolnick2025accurate}. A popular decoding procedure associated with SCF-based representations is the Yeong-Torquato (YT) method, in which the construction is formulated as an energy minimization problem \cite{Ye98a, Ye98b}, subsequently solved using simulated annealing \cite{kirkpatrick1983optimization}. The YT method exhibits superior convergence performance for binary microstructure constructions compared to, e.g., gradient-based method \cite{cheng2022data}, albeit with relatively high computational cost.

In this work, we focus on disordered hyperuniform (DHU) heterogeneous materials (see Sec. II for detailed definitions). Such materials possess a structure that is similar to liquids or glasses in that they are statistically isotropic and lack conventional long-range order, and yet they completely suppress large-scale normalized density fluctuations like crystals \cite{To03, Za09, To16a, To18a}. In this sense, disordered hyperuniform materials can be considered to possess a hidden long-range order. This unique combination of local disorder and long-range hidden order endows DHU materials with many unusual physical properties, including wave propagation characteristics \cite{ref31, ref32, ref33, scattering, granchi2022near, park2021hearing, klatt2022wave, tavakoli2022over, cheron2022wave, yu2021engineered, li2018biological}, thermal, electrical and diffusive transport properties \cite{ref34, torquato2021diffusion, maher2022characterization}, mechanical properties \cite{ref35, puig2022anisotropic} as well as optimal multifunctional characteristics \cite{ref36, kim2020multifunctional, torquato2022extraordinary}, offering many potential engineering applications. We note that hyperuniformity has been discovered in a variety of physical \cite{ref4, ref5, ref6, ref7, ref16, ref17, ref18, ref19, ref20, ref21, ref22, ref23, salvalaglio2020hyperuniform, hexner2017noise, hexner2017enhanced, weijs2017mixing,
lei2019nonequilibrium, lei2019random, ref8, ref9, ref10, ref11, ref12, ref13, ref14, ref15, ref24, ref25, sanchez2023disordered}, material \cite{ref28, ref29, ref30, Ge19, sakai2022quantum, Zh20, Ch21, PhysRevB.103.224102, Zh21, nanotube, zhang2023approach, chen2021multihyperuniform, chen2025pnas} and biological \cite{ref26, ref27, ge2023hidden, liu2024universal, tang2024tunablehyper} systems. We refer the interested readers to the recent review article by Torquato \cite{To18a} for a comprehensive discussion on hyperuniform states of matter.

The preponderance previous studies of DHU heterogeneous materials focused on microstructure constructions of two-phase media \cite{Ch18a, shi2023computational, shi2025three}, modeled by a binary random field possessing a vanishing spectral density function in the zero-wavenumber limit, i.e., $\lim_{|{\bf k}|\rightarrow 0}\Tilde{\chi}_{_V}({\bf k}) = 0$. Here, we consider heterogeneous material systems possessing continuously varying local material properties $\mathcal{K}({\bf x})$ (e.g., thermal or electrical conductivity), modeled via a random field. We devise quantitative microstructure representation of the material systems based on a class of analytical spectral density function ${\tilde \chi}_{_\mathcal{K}}({k})$ associated with $\mathcal{K}({\bf x})$, possessing a power-law small-$k$ scaling behavior ${\tilde \chi}_{_\mathcal{K}}({k}) \sim k^\alpha$. By controlling the exponent $\alpha$ and using a highly efficient forward generative model, we obtain realizations of a wide spectrum of distinct material microstructures spanning from hyperuniform ($\alpha>0$) to nonhyperuniform ($\alpha=0$) to antihyperuniform ($\alpha<0$) systems. 

We subsequently carry out a comprehensive investigation of the resulting temperature field resulting from the heterogeneous local properties, including both a numerical study and a perturbation analysis to quantitatively connect the fluctuations of the local material property to the fluctuations of the resulting physical field. In the weak-contrast limit, i.e., when the fluctuations of the property are much smaller than the average value, our first-order perturbation theory reveals that the physical fields associated with Class-I hyperuniform materials (characterized by $\alpha \ge 2$) are also hyperuniform, albeit with a lower hyperuniformity exponent ($\alpha-2$). As one moves away from this weak-contrast limit, the fluctuations of the physical field develop a diverging spectral density at the origin, revealed by our higher-order analysis and verified by our numerical results. We also establish an end-to-end mapping connecting the spectral density of the local material property to the overall effective conductivity of the material system via numerical homogenization. We observe a sharp decrease of the variance of effective properties across realizations as $\alpha$ increases from antihyperuniform values to hyperuniform values. Our results have significant implications for the design of novel DHU materials with targeted physical properties.

The rest of the paper is organized as follows: In Sec. II, we provide definition of Gaussian random fields, correlation function, spectral density function, and hyperuniformity in heterogeneous material systems, as well as the effective material properties of interest. In Sec. III, we present the microstructure representation framework and constructions results of a wide spectrum of hyperuniform and nonhyperuniform realizations with prescribed analytical spectral density functions. In Sec. IV, we present numerical and analytical results on the temperature field fluctuations. In Sec. V, we present the results on effective material properties. In Sec. VI, we provide concluding remarks and outlook of future work.

\section{Definitions and Preliminaries}
%\section{Correlation Function, Spectral Density, and Hyperuniformity of Heterogeneous Materials}
\label{definition}

%In our subsequent discussion, the analytical 
%We can briefly introduce C2 here, for subsquent quantification of reconstruction patterns.

\subsection{Gaussian Random Fields and Correlation Functions}

A \emph{Gaussian random field} (GRF) is a stochastic process \(I(\mathbf{x})\) defined on a continuous domain (e.g., \(\mathbf{x} \in \mathbb{R}^d\)) such that for any finite collection of points \(\mathbf{x}_1,\mathbf{x}_2,\dots,\mathbf{x}_N\), the vector
\begin{equation}
\left(I(\mathbf{x}_1), I(\mathbf{x}_2), \dots, I(\mathbf{x}_N)\right)
\end{equation}
follows a multivariate Gaussian distribution \cite{AdlerTaylor2007}. Consequently, a GRF is completely characterized by its mean 
\begin{equation}
\phi(\mathbf{x}) = \mathbb{E}[I(\mathbf{x})],
\end{equation}
and its covariance function
\begin{equation}
C(\mathbf{x}_1,\mathbf{x}_2) = \mathbb{E}\Bigl[(I(\mathbf{x}_1)-\phi(\mathbf{x}_1))(I(\mathbf{x}_2)-\phi(\mathbf{x}_2))\Bigr].
\end{equation}
For a stationary GRF, \(\phi(\mathbf{x}) = \phi\) is position-independent constant and the covariance function depends only on the displacement, i.e., \(C(\mathbf{x}_1,\mathbf{x}_2)=C(\mathbf{r})\) with \(\mathbf{r} = \mathbf{x}_2-\mathbf{x}_1\). In the case of an isotropic GRF, \(C(\mathbf{r}) = C(r)\) depends only on the Euclidean distance \(r = |\mathbf{r}|\). For a GRF without long-range correlations, $C({\bf r})$ possesses the following asymptotic behavior:
\begin{equation}
\lim_{|{\bf r}|\rightarrow \infty} C(\mathbf{r}) = 0.
\end{equation}

According to Bochner's theorem, the covariance function of a stationary process is the Fourier transform of a nonnegative measure. When this measure has a density, the \emph{power spectral density} (PSD) is defined as
\begin{equation}
\tilde{\chi}(\mathbf{k}) = \int_{\mathbb{R}^n} e^{-i\mathbf{k}\cdot\mathbf{r}}\,C(\mathbf{r})\,d\mathbf{r},
\end{equation}
where \(\tilde{\chi}(\mathbf{k})\) is nonnegative and symmetric, i.e., \(\tilde{\chi}(-\mathbf{k}) = \tilde{\chi}(\mathbf{k})\). In many applications, the small-\(k\) scaling of \(\tilde{\chi}(\mathbf{k})\) (e.g., \(\tilde{\chi}(\mathbf{k}) \sim |{\bf k}|^\alpha\)) plays a critical role in determining the large-scale correlations of the field, a property central to the concept of hyperuniformity as discussed in detail below. 
%Moreover, the spatial structure of the microstructure is also characterized by higher-order, \(n\)-point correlation functions \(S^{(i)}_n\); in many cases, the two-point correlation function \(S^{(i)}_2\) captures the essential features of the field.

GRFs have been widely employed to model heterogeneous material systems \cite{torquato2002random, Sa03a, jeulin2021morphological, roberts1997statistical}. For example, in the case of a binary alloy with a stable solid solution phase, the spatial fluctuations of element concentrations can lead to spatial fluctuations in local material properties such as electrical/thermal conductivity and elastic moduli, which can be very well modeled using GRFs. By specifying the covariance function (or equivalently the spectral density \(\tilde{\chi}(\mathbf{k})\)), one can control the correlations of the property fluctuations across scales, offering an effective approach for the design and engineering of such material systems to achieve targeted overall material properties and performance. Realizations of the GRFs associated with specific \(\tilde{\chi}(\mathbf{k})\) correspond to the representative volume elements of the material system with varying local properties for subsequent numerical analysis.

%In particular, for disordered hyperuniform materials, one designs \(\tilde{\chi}(\mathbf{k})\) so that it vanishes as \(|\mathbf{k}|\to 0\) (often with a power-law scaling), thereby suppressing long-range density fluctuations \cite{TorquatoStillinger2003}. This property enables the generation of microstructures with highly controlled long-range order despite local disorder.

\subsection{Hyperuniform, Nonhyperuniform and Antihyperuniform Random Field}

%.The hyperuniformity concept was first introduced for point configurations \cite{To03, To18a} and was subsequently generalized to binary heterogeneous materials \cite{Za09} and random scalar, vector and tensor fields \cite{To16a}. A statistically homogeneous random medium is specified by the indicator function $I({\bf x})$ of the reference phase, which is also a binary random field.

In the context of a scalar random field, the quantity of interest is the local field variance $\sigma_{_F}^2(R)$ \cite{To16a, Ma17a}:
\begin{equation}
\sigma_{_F}^2(R) = \frac{1}{v_1(R)}\int_{\mathbb{R}^d} I({\bf r})\alpha_2(r; R)d{\bf r},
\label{eq_field_fluc}
\end{equation}
where $\alpha_2(r; R)$ is the scaled intersection volume, i.e., the intersection volume of two spherical windows of radius $R$ whose centers are separated by a distance $r$, divided by the volume $v_1(R)$ of the window, i.e.,
\begin{equation}
v_1(R)=\frac{\pi^{d/2} R^d}{\Gamma(1+d/2)}.
\label{v1}
\end{equation}

%defined as the integration of indicator function within the observation window over the window volume, which fluctuates as the window randomly moves in the system. Similar to the case of point configurations, a spherical observation window with radius $R$ and volume

%\begin{equation}
%v_1(R)=\frac{\pi^{d/2} R^d}{\Gamma(1+d/2)}
%\label{v1}
%\end{equation}
%is employed. The associated variance $\sigma_{_V}^2(R)$ is given by

%To03, Za09, To16a, To18a

A disordered hyperuniform random field is one whose
$\sigma_{_F}^2(R)$ decreases more rapidly than $R^d$ for large $R$ \cite{To16a}, i.e.,
\begin{equation}
\lim_{R\rightarrow\infty}\sigma_{_F}^2(R) \cdot R^d = 0.
\end{equation}
This behavior is to be contrasted with those of typical random field for which the variance decays as $R^{-d}$, i.e., as the inverse of the window volume $v_1(R)$.

%Equivalently, the hyperuniformity of a random medium can be defined via its autocovariance function $\chi_{_V}({\bf r})$.

The hyperuniform condition is equivalently given by
\begin{equation}
\label{eq_hyper} \lim_{|{\bf k}|\rightarrow 0}\Tilde{\chi}({\bf k})
= 0,
\end{equation}
which implies that  the direct-space autocovariance function $C({\bf r})$ exhibits both positive  and  negative correlations such that its volume integral over all space is exactly zero \cite{To16b}, i.e.,
\begin{equation}
\int_{\mathbb{R}^d} C({\bf r})d{\bf r} = 0.
\label{eq_sum_rule}
\end{equation}
Eq. (\ref{eq_sum_rule}) is a direct-space sum rule for hyperuniformity of random fields. 

%We note that the spectral density function $\Tilde{\chi}_{_V}({\bf k})$ characterizes the light scattering property of the material, and its value is proportional to the scattering intensity of an incident wave of wavevector ${\bf k}$ \cite{ishimaru1978wave}. This allows us to directly engineer scattering properties of heterogeneous materials by imposing a desirable $\Tilde{\chi}_{_V}({\bf k})$.

For hyperuniform random fields whose spectral density goes to zero as a power-law scaling as $|\bf k|$ tends to zero \cite{To16a}, i.e.,
\begin{equation}
    {\tilde \chi}({\bf k})\sim |{\bf k}|^\alpha,
\end{equation}
the small-$k$ of ${\tilde \chi}({\bf k})$ determines the large-$R$ behavior of the variance $\sigma_{_F}^2(R)$. There are three different scaling regimes (classes) that describe the associated large-$R$ behaviors of the local volume fraction variance:
\begin{equation}
\sigma^2_{_F}(R) \sim
\begin{cases}
R^{-(d+1)}, \quad\quad\quad \alpha >1 \qquad &\text{(Class I)}\\
R^{-(d+1)} \ln R, \quad \alpha = 1 \qquad &\text{(Class II)}\\
R^{-(d+\alpha)}, \quad 0 < \alpha < 1\qquad  &\text{(Class III).}
\end{cases}
\label{eq:classes}
\end{equation}
Classes I and III are the strongest and weakest forms of hyperuniformity, respectively. Class I systems include all crystal structures \cite{To03}, many quasicrystal structures \cite{Og17} and exotic disordered media \cite{Za09, Ch18a}. Examples of Class II systems include some quasicrystal structures \cite{Og17}, perfect glasses \cite{zhang2017classical}, and maximally random jammed packings \cite{ref4, ref5, ref6, Za11c, Za11d}. Examples of Class III systems include classical disordered ground states \cite{Za11b}, random organization models \cite{ref20}, perfect glasses \cite{zhang2017classical}, and perturbed lattices \cite{Ki18a}; see Ref. \cite{To18a} for a more comprehensive list of systems that fall into the three hyperuniformity classes. 

Stealthy hyperuniform fields are a special subset of Class III systems in which the spectral density function is exactly zero for a range of wavevectors around the origin \cite{To18a, torquato2021diffusion}, i.e.,
\begin{equation}
    {\tilde \chi}({\bf k}) = 0, \quad {\bf k} \in \Omega,
\end{equation}
where $\Omega$ is a finite region around the origin of the Fourier space. Stealthy systems can be approximately considered to possess a power-law spectral density in the infinite-$\alpha$ limit, i.e., ${\tilde \chi}({\bf k})\sim |{\bf k}|^\alpha$ with $\alpha \rightarrow \infty$.

%\textcolor{red}{Need to briefly define standard nonhyperuniform and antihyperuniform systems.}
%\textcolor{blue}{Accordingly, the local volume fraction variance has the following large-$R$ scaling behavior:
%\begin{equation}
%\sigma^2_{_V}(R) \sim R^{-d}, \quad \alpha = 0.
%\end{equation}}

By contrast, for any nonhyperuniform random fields, the local field variance has the following large-$R$ scaling behaviors \cite{torquato2021diffusion}:
\begin{equation}
\sigma^2_{_F}(R) \sim
\begin{cases}
R^{-d}, ~\alpha =0 \quad \text{(standard nonhyperuniform)}\\
R^{-(d+\alpha)}, ~-d < \alpha < 0\quad  \text{(antihyperuniform).}
\end{cases}
\label{eq:classesnon}
\end{equation}
A {\it standard nonhyperuniform} random field \cite{torquato2021diffusion} is one whose spectral density function is bounded and approaches a finite constant as $|{\bf k}|$ goes to zero, i.e., 
\begin{equation}
   \lim_{|{\bf k}|\rightarrow 0} {\tilde \chi}({\bf k})\sim const.
\end{equation}
Examples of standard nonhyperuniform systems include overlapping systems with Poisson distribution of centers, equilibrium hard-sphere fluids, and hard-sphere packings generated via random sequential addition process \cite{To18a, torquato2021local}. An {\it antihyperuniform} random field \cite{torquato2021diffusion} possesses an unbounded spectral density function in the zero-$|{\bf k}|$ limit, i.e., 
\begin{equation}
   \lim_{|{\bf k}|\rightarrow 0} {\tilde \chi}({\bf k})\rightarrow +\infty
\end{equation}
Systems at critical point containing macroscopic scale fluctuations possess a diverging spectral density at the origin and thus are antihyperuniform. Other examples include systems generated via hyperplane intersection process (HIP) and Poisson cluster process \cite{torquato2021local}.

\subsection{Effective Conductivity in Heterogeneous Materials}

Consider a heterogeneous material in a bounded domain \(\Omega\subset \mathbb{R}^d\) with a spatially varying locally isotropic conductivity \(\mathcal{K}(\mathbf{x})\). The steady‐state conduction in such a medium is governed by the partial differential equation (PDE)
\begin{equation}
-\nabla\cdot\Bigl(\sigma(\mathbf{x})\nabla T(\mathbf{x})\Bigr)=0 \quad \text{in } \Omega,
\label{eq_heat_eq}
\end{equation}
subject to appropriate boundary conditions (e.g., periodic boundary conditions), where \(T(\mathbf{x})\) represents the temperature (or electric potential) field. We note that although our subsequent analysis explicitly considers the heat conduction problem, by mathematical analogy, all of the analysis and results immediately apply to electric conduction as well.

In order to characterize the macroscopic transport properties of the medium, the local flux is defined by
\begin{equation}
\mathbf{J}(\mathbf{x}) = -\sigma(\mathbf{x})\nabla T(\mathbf{x}),
\end{equation}
and the effective conductivity tensor
\begin{equation}
\boldsymbol{\Sigma}^* = \begin{bmatrix}
\mathcal{K}^e_{xx} & \mathcal{K}^e_{xy}  \\
\mathcal{K}^e_{xy}  & \mathcal{K}^e_{yy} 
\end{bmatrix}
\end{equation}
is defined such that the volume-averaged flux equals the response of an equivalent homogeneous medium under the same macroscopic gradient:
\begin{equation}
\langle \mathbf{J}(\mathbf{x}) \rangle = -\boldsymbol{\Sigma}^*\langle \nabla T(\mathbf{x}) \rangle,
\end{equation}
where \(\langle \cdot \rangle\) denotes the spatial (or ensemble) average over \(\Omega\). For a prescribed macroscopic gradient \(\mathbf{G}\), we thus have
\begin{equation}
\boldsymbol{\Sigma}^*\,\mathbf{G} = -\langle \sigma(\mathbf{x})\nabla T(\mathbf{x})\rangle.
\end{equation}
For statistically isotropic materials, which is the main focus of this work, the off-diagonal component $K_{xy}^e = 0$. In the weak-contrast case, when for anisotropic systems (e.g., certain antihyperuniform materials constructed below), one still has $K_{xy}^e \ll K_{xx}^{e}$. Therefore, we will focus on the principal components of the effective conductivity tensor in subsquent studies. In particular, if the applied unitary gradient is \(\mathbf{G}=(1,0)\), the effective conductivity in the \(x\)-direction is defined as
\begin{equation}
\mathcal{K}^e_{xx} = -\langle J_x(\mathbf{x})\rangle,
\label{eq_Kexx}
\end{equation}
and similarly for \(\mathbf{G}=(0,1)\), one obtains
\begin{equation}
\mathcal{K}^e_{yy} = -\langle J_y(\mathbf{x})\rangle.
\label{eq_Keyy}
\end{equation}

%the temperature field is decomposed as
%\[
%T(\mathbf{x}) = T_{\mathrm{macro}}(\mathbf{x}) + T_p(\mathbf{x}),
%\]
%where
%\[
%T_{\mathrm{macro}}(\mathbf{x}) = \text{offset} + \mathbf{G}\cdot\mathbf{x},
%\]
%with \(\mathbf{G}\) being the imposed constant macroscopic gradient, and \(T_p(\mathbf{x})\) is a periodic perturbation field that accounts for the local fluctuations induced by the heterogeneity.

A common approach to compute \(\boldsymbol{\Sigma}^*\) is to solve the so-called \emph{cell problem}. That is, one seeks a periodic function \(T_p(\mathbf{x})\) (often with a constraint to eliminate the constant nullspace) such that
\begin{equation}
\nabla\cdot\Bigl(\sigma(\mathbf{x})\bigl(\mathbf{G}+\nabla T_p(\mathbf{x})\bigr)\Bigr)=0 \quad \text{in } \Omega,
\end{equation}
with \(T_p(\mathbf{x})\) being periodic over \(\Omega\). Once the solution is obtained, the effective conductivity is computed from the volume average of the local flux:
\begin{equation}
\boldsymbol{\Sigma}^* = -\frac{1}{|\Omega|} \int_{\Omega} \sigma(\mathbf{x})\Bigl[\mathbf{G}+\nabla T_p(\mathbf{x})\Bigr]\,d\mathbf{x}.
\end{equation}
This definition ensures that the heterogeneous medium exhibits the same macroscopic response as an equivalent homogeneous medium with conductivity \(\boldsymbol{\Sigma}^*\) when subject to the same driving field \cite{Bensoussan1978,Torquato2002}.

The effective conductivity $\boldsymbol{\Sigma}^*$ captures the macroscopic transport properties of a heterogeneous material by averaging the local response of the microstructure. Its precise definition, based on the solution of the cell problem and the associated variational formulation, forms a crucial link between the microstructural disorder (characterized by, e.g., the spectral density \(\tilde{\chi}(\mathbf{k})\)) and the overall macroscopic behavior of the material.

\section{Microstructure Representation and Construction}

In this section, we present a microstructure representation framework for disorder heterogeneous materials systems via analytical spectral density \(\tilde{\chi}_{_\mathcal{K}}(\mathbf{k})\), which characterizes the fluctuations of the local conductivity $\mathcal{K}({\bf x})$. As shown below, by controlling the long-range correlations of the system via the ``hyperuniformity exponent'' $\alpha$, we can obtain a wide spectrum of distinct materials spanning from antihyperuniform to standard nonhyperuniform to hyperuniform systems. Realizations of these distinct materials are constructed using a highly efficient generative model based on the targeted \(\tilde{\chi}_{_\mathcal{K}}(\mathbf{k})\).

%description of the microstructure in disordered hyperuniform materials. We first introduce the spectral density function for local conductivity fluctuations, explain the role of each parameter, and then describe the generative model that renders realizations from a white noise field. Finally, we discuss numerical results showing the evolution of key structural features as the parameter \(\alpha\) is varied from positive (stealthy hyperuniform) to negative (antihyperuniform) values.

\begin{figure*}[htbp]
  \centering
  \includegraphics[width=0.98\textwidth]{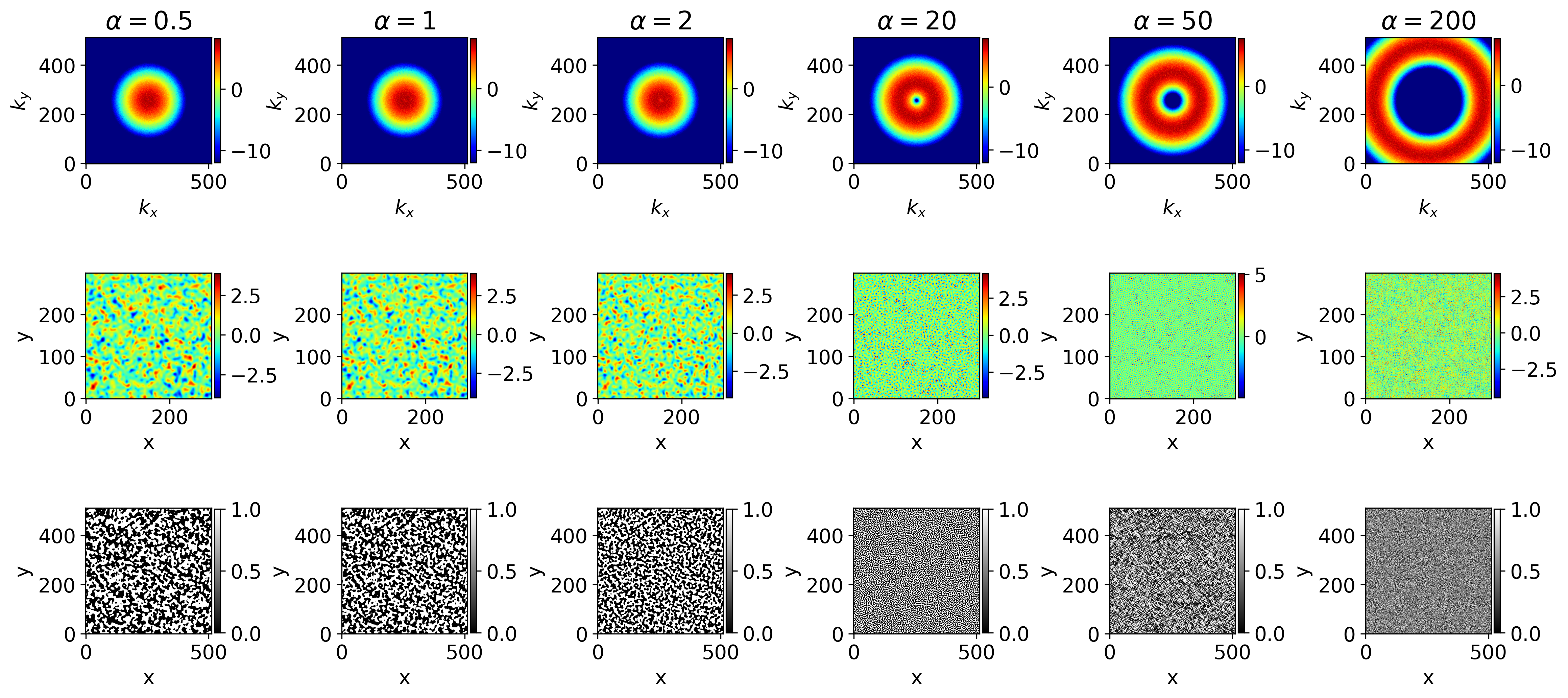}
  \caption{Realizations of the local conductivity field (middle panels) associated with the analytical spectral density function \(\tilde{\chi}_{_\mathcal{K}}(\mathbf{k})\) in log scale (upper panels). Lower panels show binarized fields (i.e., $\delta \mathcal{K}({\bf x})>0$ shown as white) for better visualization of the morphological features. From left to right, $\alpha = 1/2$, 1, 2, 20, 50, 200, corresponding to all classes of hyperuniformity.}
  \label{fig:composite1}
\end{figure*}
%Composite of real-space fields (left) and log power spectra (right) for  \(\alpha \in \{\,\tfrac{1}{2},\,1,\,2,\,20,\,50,\,200\}\). As \(\alpha\) increases from \(1/2\) to large positive values, the suppression of long-wavelength modes (low \(k\)) becomes increasingly pronounced, reflecting a transition toward hyperuniform behavior.

\begin{figure*}[htbp]
  \centering
  \includegraphics[width=0.98\textwidth]{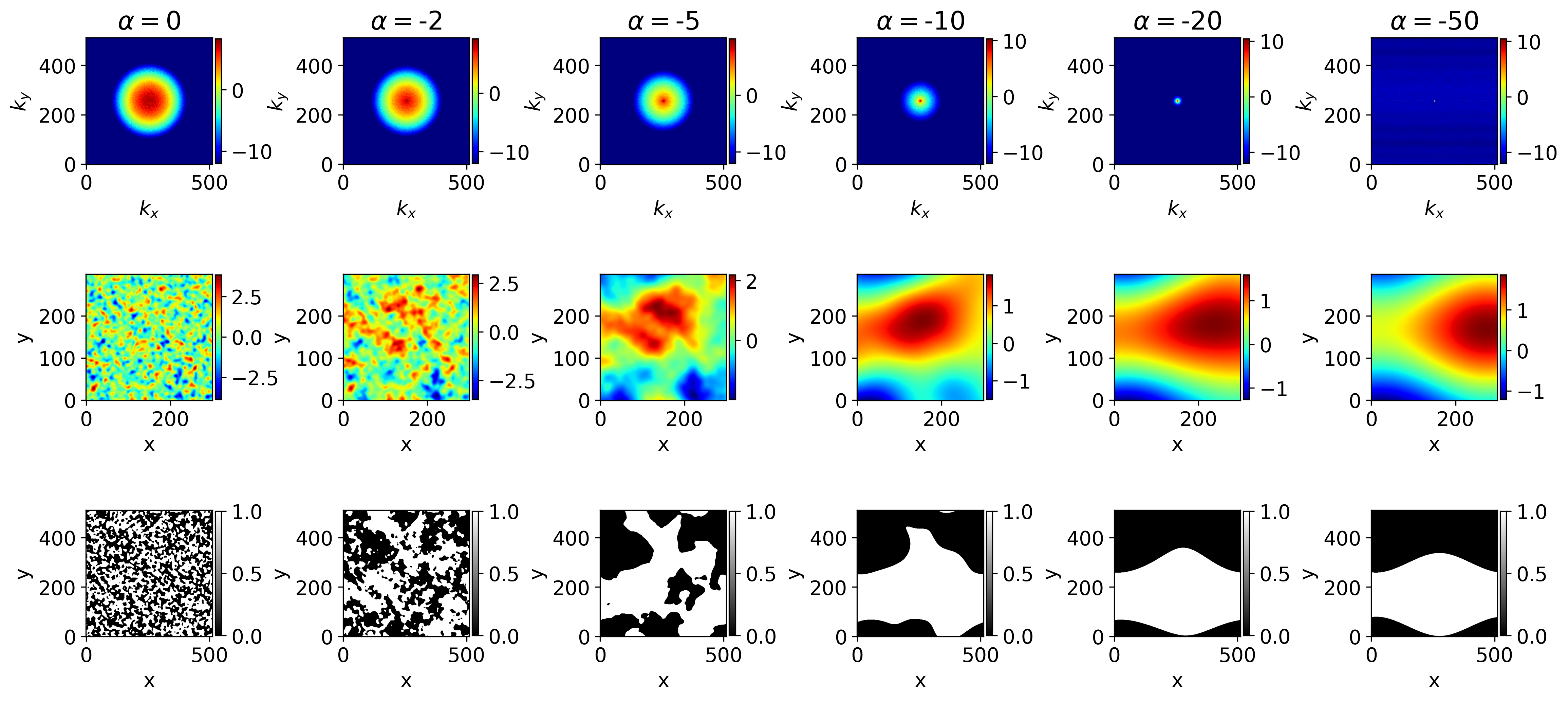}
  \caption{Realizations of the local conductivity field (middle panels) associated with the analytical spectral density function in log scale \(\tilde{\chi}_{_\mathcal{K}}(\mathbf{k})\) (upper panels). Lower panels show binarized fields (i.e., $\delta \mathcal{K}({\bf x})>0$ shown as white) for better visualization of the morphological features. From left to right, $\alpha = 0$, -2, -5, -10, -20, -50, corresponding to nonhyperuniform and antihyperuniform systems. }
  \label{fig:composite2}
\end{figure*}

%Composite of real-space fields (left) and log power spectra (right) for \(\alpha \in \{\,0,\,-2,\,-5,\,-10,\,-20,\,-50\}\). As \(\alpha\) moves into the negative range, low-frequency (long-wavelength) modes become progressively amplified, leading to large-scale inhomogeneities characteristic of antihyperuniform systems.

\subsection{Microstructure Representation via Analytical Spectral Density Function for Local Conductivity}

%that characterizes the spatial frequency content of local conductivity fluctuations. In our model, the spectral density is defined by

%\\
%(k + k_0)^{\alpha}\exp\!\Bigl(-\frac{k^2}{2\sigma^2}\Bigr), \quad \alpha <0

%\(k_0\) is a small positive constant that prevents divergence as \(k\to 0\) and sets a lower cutoff for the spectral density.

Here we define the {\it microstructure} of a heterogeneous material as the scalar field characterizing its locally isotropic conductivity $\mathcal{K}({\bf x})$, which contains a constant part and a fluctuating part, i.e., 
\begin{equation}
\mathcal{K}({\bf x}) = \mathcal{K}_0 + \delta \mathcal{K}({\bf x})
\end{equation}
and is associated with an analytical spectral density function \(\tilde{\chi}_{_\mathcal{K}}({k})\): 
\begin{equation} \label{eq:chi}
  \tilde{\chi}_{_\mathcal{K}}({k}) = 
 \begin{cases}
k^{\alpha}\exp\!\Bigl(-\frac{k^2}{2\sigma^2}\Bigr), \quad \alpha \neq 0 \\
 \exp\!\Bigl(-\frac{k^2}{2\sigma^2}\Bigr), \quad  \alpha = 0 
\end{cases} 
\end{equation}
where \(k = \|\mathbf{k}\|\) is the magnitude of the wavevector; \(\alpha\) is a dimensionless exponent that governs the small-$k$ scaling behavior of \(\tilde{\chi}_{_\mathcal{K}}({k})\) and thus the large-scale fluctuations of $\mathcal{K}({\bf x})$; one obtains hyperuniform, nonhyperuniform and antihyperuniform microstructures respectively for $\alpha >0$,  $\alpha =0$ and $\alpha <0$; \(\sigma\) controls the exponential decay at high $k$ values and thus, determines the ``smoothness'' of the field at small scales.

%The parameters are defined as follows:
%\begin{itemize}
%  \item \(k_0\) is a small regularization parameter that prevents divergence as \(k\to 0\) and sets a lower cutoff for the spectral density.
%  \item \(\sigma\) controls the exponential decay at high frequencies, thereby determining the smoothness of the field at small scales.
%  \item \(\alpha\) is a dimensionless exponent that governs the low-frequency behavior. For \(\alpha = 0\), the spectral density reduces to
%  \[
%  \tilde{\chi}(\mathbf{k}) = \exp\!\Bigl(-\frac{k^2}{2\sigma^2}\Bigr),
%  \]
%  which decays in a Gaussian manner.  For \(\alpha > 0\), low-wavenumber components are suppressed (\(\tilde{\chi}(k) \to 0\) as \(k \to 0\)), a signature of hyperuniformity. In contrast, for \(\alpha < 0\) the spectral weight is enhanced at low \(k\), indicating the presence of strong long-range fluctuations (antihyperuniform behavior).
%\end{itemize}

In the subsequent discussions, we will focus on the ``hyperuniformity'' exponent $\alpha$ as the tuning knob for obtaining distinct microstructures. For \(\alpha > 0\), the low-wavenumber components are suppressed, i.e., \(\tilde{\chi}(k)_{_\mathcal{K}} \to 0\) as \(k \to 0\), a signature of hyperuniformity. In contrast, for \(\alpha < 0\) the spectral weight is enhanced at low \(k\), indicating the presence of strong long-range fluctuations which is typical antihyperuniform behavior. Thus, by tuning \(\alpha\), we can model a continuum spectral of distinct microstructures: from stealthy-hyperuniform-like systems associated with high positive \(\alpha\), to a standard Gaussian system with \(\alpha = 0\), to structures with enhanced large-scale heterogeneity with negative \(\alpha\). 

\subsection{Generative Model: From White Noise to Structured Field }

Our generative model employs a Fourier filtering approach to produce realizations of the local conductivity field with the target spectral density given in Eq.~\eqref{eq:chi}. In particular, an initial white noise field is iteratively transformed into a structured conductivity field via frequency-domain shaping \cite{shinozuka1991simulation, woodchan1994}. The process consists of the following steps:

\begin{enumerate}
  \item \textbf{Initialization:} A white noise field \(w(\mathbf{x})\) is generated on a discrete \(N\times N\) grid over a domain of size \(L\times L\). Each grid point is assigned an independent Gaussian random value with zero mean and unit variance.

  \item \textbf{Fourier Transform:} The discrete Fourier transform (DFT) of the white noise field, \(\hat{w}(\mathbf{k}) = \mathcal{F}\{w(\mathbf{x})\}\), is computed. Since \(w(\mathbf{x})\) is real valued, the Fourier coefficients satisfy Hermitian symmetry.

  \item \textbf{Spectral Filtering:} The Fourier coefficients are modified by rescaling them with the square-root of the desired spectral density. Specifically, we define
  \begin{equation} \label{eq:filter}
    \hat{u}(\mathbf{k}) = \sqrt{\tilde{\chi}_{_\mathcal{K}}(\mathbf{k})}\,\hat{w}(\mathbf{k}),
  \end{equation}
  so that
  \begin{equation}
  |\hat{u}(\mathbf{k})|^2 \propto \tilde{\chi}_{_\mathcal{K}}(\mathbf{k}).
  \end{equation}
  The random phases from \(\hat{w}(\mathbf{k})\) are preserved to maintain randomness in the field. This step imposes the prescribed two-point statistics (i.e., $C(r)$) on the field.

  \item \textbf{Inverse Transform:} The inverse Fourier transform is performed to obtain the real-space field:
  \begin{equation}
  u(\mathbf{x}) = \mathcal{F}^{-1}\{\hat{u}(\mathbf{k})\}.
  \end{equation}
  The result is a continuous Gaussian random field that exhibits the desired spectral density \(\tilde{\chi}(\mathbf{k})\). Because the filtering is a linear operation, the field remains Gaussian-distributed.

  \item \textbf{Normalization and Constraints:} Finally, we impose constraints on the field by subtracting its spatial average (to enforce zero mean) and rescaling it so that its standard deviation equals a target value (e.g., unity). This normalization ensures that differences among realizations stem solely from the imposed microstructural correlations, not from trivial shifts or scaling.
\end{enumerate}

 %By tuning the parameter \(\alpha\) , one can control the suppression or enhancement of low-frequency fluctuations, thereby generating a wide range of microstructures from stealthy hyperuniform (for \(\alpha>0\)) to antihyperuniform (for \(\alpha<0\)).

\subsection{Construction Results and Structural Evolution}

We employ the Fourier filtering method to generate realizations of the conductivity fields associated with the analytical spectral density function \(\tilde{\chi}_{_\mathcal{K}}(\mathbf{k})\) given by Eq. (\ref{eq:chi}). Using a Fast Fourier Transform (FFT) implement, the method scales as \(O(N \log N)\), where \(N\) is the grid resolution per spatial dimension. 
In our simulations, we consider a two-dimensional domain of size \(L = 50\) with a grid resolution of \(N = 512\), resulting in a grid spacing of \(dx = L/N\). For each realization, the generation time is approximately \(6\)–\(7\) milliseconds. For a \(512\times512\) grid in double precision, the required storage is on the order of a few megabytes. This high efficiency enables rapid generation of large ensembles in parallel on high-performance computing clusters, which is essential for statistical studies of effective material properties. For the antihyperuniform systems, we add a small positive off-set $k_0 = 0.1$ to avoid numerical divergence, which does not affect the underlying physics.

%The spectral density function incorporates a regularization parameter \(k_0 = 0.1\) to prevent singularities when \(\alpha < 0\) and to impose a lower bound for the spectral content.

%The computational cost of generating a single microstructure is highly efficient. 

%Memory usage is moderate;  Additionally, since each realization is generated independently, the algorithm is inherently parallelizable, making it well-suited for deployment on high-performance computing clusters.

%This setup ensures that the generated microstructures capture the desired statistical properties while maintaining computational efficiency and scalability for large-scale simulations.

%Our numerical simulations reveal a clear evolution of the microstructural features and their corresponding power spectra as the hyperuniform exponent \(\alpha\) varies. To illustrate this progression, we combine multiple \(\alpha\) values into two composite figures. In each figure, the \emph{left panel} of each pair displays the real-space conductivity field (typically cropped to highlight structural details), and the \emph{right panel} of each pair shows the corresponding logarithm of the power spectral density. The top rows in each composite correspond to larger \(\alpha\), while the bottom rows correspond to smaller (or more negative) \(\alpha\). 

Figure~\ref{fig:composite1} shows the numerical construction results for hyperuniform systems ($\alpha > 0$), including Class-III ($\alpha = 1/2$), Class-II ($\alpha = 1$) and Class-I ($\alpha \ge 2$) systems. In particular, the targeted spectral density functions \(\tilde{\chi}_{_\mathcal{K}}(\mathbf{k})\) are shown in the upper panels, and the associated realizations of the local conductivity field are shown in the middle panels. The lower panels show the binarized fields (i.e., $\delta \mathcal{K}({\bf x})>0$ shown as white) for better visualization of the morphological features. It can be seen the hyperuniform fields contains features with well-defined size and morphology, resulting in an overall much uniform distribution. As $\alpha$ increases, the size of the features decreases, leading to finer and finer structures. The constructed field also develops a labyrinth pattern with well-defined uniform wavelengths for very large $\alpha$ (e.g., $\ge 20$). These results are consistent with previous studies of standard and stealthy hyperuniform binary fields \cite{chen2018designing, shi2023computational}. Indeed, it can be clearly seen that \(\tilde{\chi}_{_\mathcal{K}}(\mathbf{k})\) with a large $\alpha$ exhibits a clear ``exclusion'' region mimicking that of a stealthy hyperuniform system.

Figure~\ref{fig:composite2} shows the numerical construction results for nonhyperuniform ($\alpha = 0$) and antihyperuniform ($\alpha < 0$) systems, including the targeted spectral density functions \(\tilde{\chi}_{_\mathcal{K}}(\mathbf{k})\) (upper panels), the realizations of the local conductivity (middle panels), and the associated binarized fields (lower panels) for better visualization of the morphological features. It can be clearly seen that the realizations of the conductivity fields contain large clustering regions of positive or negative fluctuations, which significantly increase in size as $\alpha$ decreases. Such clustering contributes to the large field fluctuations within the sampling window (see Eq. (\ref{eq_field_fluc})), which is the hallmark of nonhyperuniform and antihyperuniform behaviors. For very negative $\alpha$ values, the realizations exhibit ``phase separate'' regions of high and low conductivity values, mimicking a system at critical points, characterized by a diverging spectral density at the zero-wavenumber limit. These results are consistent with previous constructions of nonhyperuniform and antihyperuniform realizations of binary fields \cite{shi2025three}. 

In summary, our spectral-density targeted generative model offers a flexible and efficient means to produce synthetic microstructures with prescribed two-point statistics. By tuning the hyperuniformity exponent \(\alpha\), one can design a wide range of materials, i.e., from stealthy hyperuniform with suppressed large-scale fluctuations to antihyperuniform with pronounced macroscopic inhomogeneities. When \(\alpha\) is large and positive, the resulting system is stealthy hyperuniform, which would essentially eliminate macroscopic variations in conductivity, as shown below. As \(\alpha\) moves toward negative values, the conductivity field becomes dominated by broad, large-wavelength fluctuations, forming highly inhomogeneous domains characteristic of antihyperuniform media. This capability to transition between such extremes highlights the versatility of spectral-based generative models for microstructure design.

\begin{figure*}[htbp]
    \centering
    \includegraphics[width=1\textwidth]{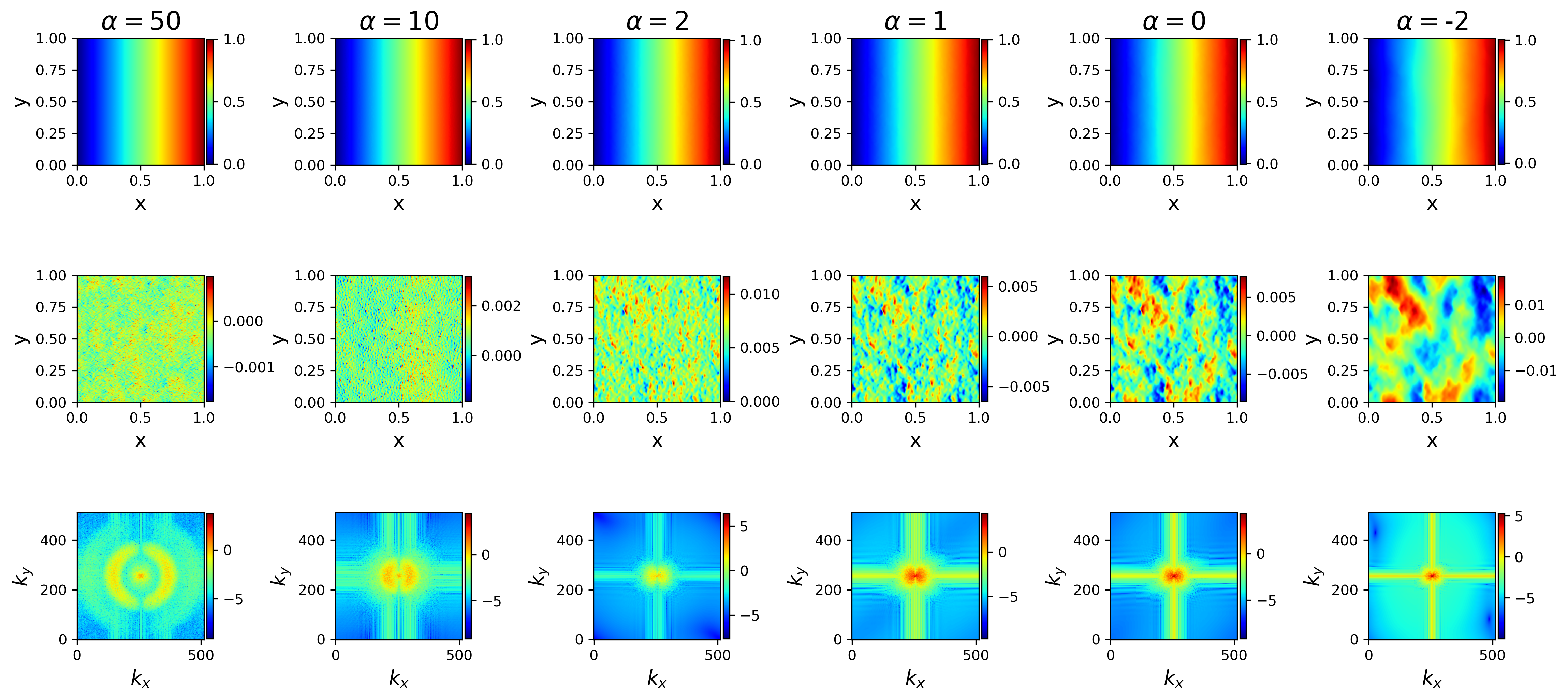}
    \caption{Steady-state temperature field (upper panels) $T({\bf x})$, temperate fluctuations (middle panels) $T_p({\bf x})$, and the associated spectral density functions \(\tilde{\chi}_{_\mathcal{T}}(\mathbf{k})\) (lower panels) for various $\alpha$ values, resulted from unitary applied gradients ${\bf G} = (1, 0)$ along the horizontal direction.
    }
    \label{fig:1}
\end{figure*}
%Temperature field perturbations for different values of $\alpha$ under a macroscopic gradient of $(1,\,0)$ applied in the $x$-direction. Lower $\alpha$ (stronger long-range conductivity fluctuations) leads to more prominent large-scale temperature variations.

\begin{figure*}[htbp]
    \centering
    \includegraphics[width=1\textwidth]{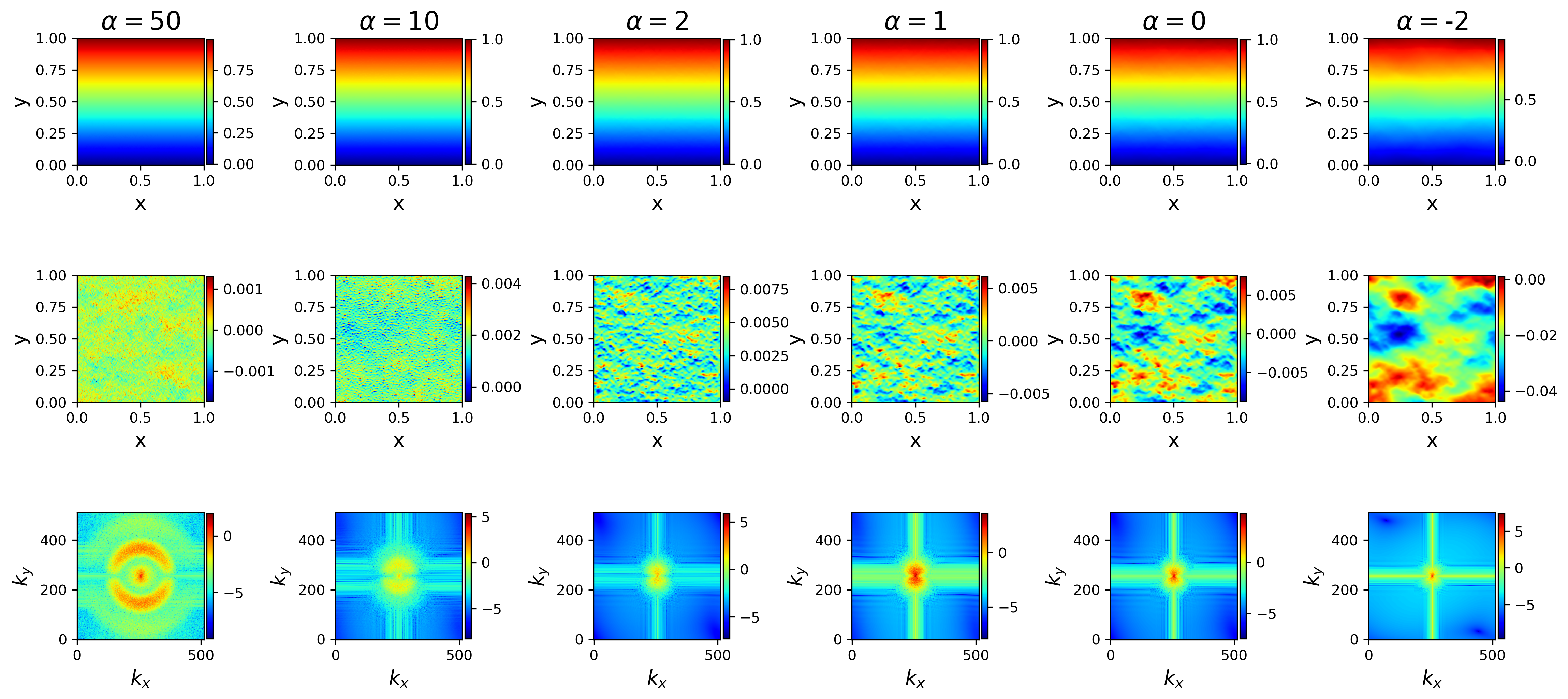}
    \caption{Steady-state temperature field (upper panels) $T({\bf x})$, temperate fluctuations (middle panels) $T_p({\bf x})$, and the associated spectral density functions \(\tilde{\chi}_{_\mathcal{T}}(\mathbf{k})\) (lower panels) for various $\alpha$ values, resulted from unitary applied gradients ${\bf G} = (0, 1)$ along the vertical direction.
    }
    \label{fig:2}
\end{figure*}

\begin{figure*}[htbp]
    \centering
    \includegraphics[width=1\textwidth]{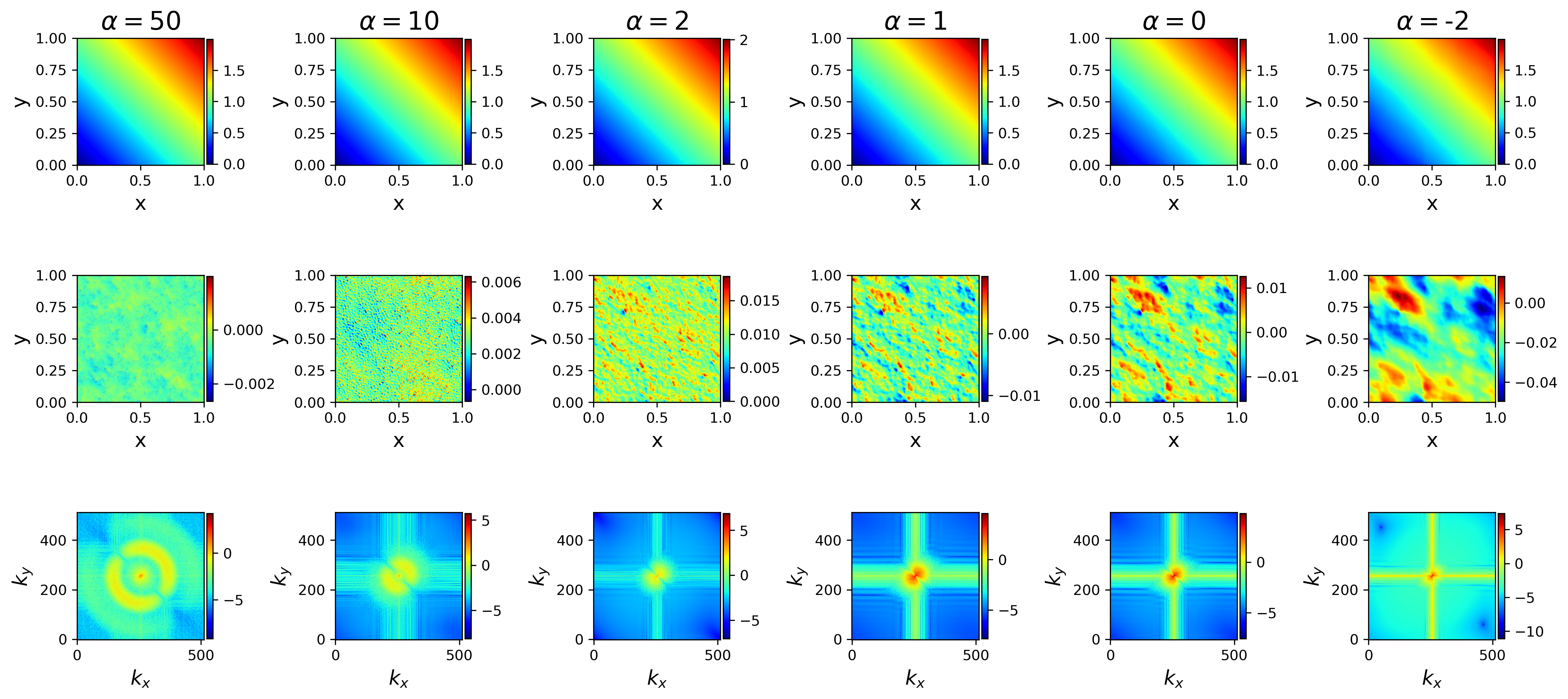}
    \caption{Steady-state temperature field (upper panels) $T({\bf x})$, temperate fluctuations (middle panels) $T_p({\bf x})$, and the associated spectral density functions \(\tilde{\chi}_{_\mathcal{T}}(\mathbf{k})\) (lower panels) for various $\alpha$ values, resulted from unitary applied gradients ${\bf G} = (1, 1)$ along the diagonal direction.
    }
    \label{fig:3}
\end{figure*}

\section{Fluctuations of Temperature Field}

In this section, we investigate the spatial fluctuations of the temperature field induced by the spatially varying local conductivity field in heterogeneous materials. Understanding the field fluctuations resulted from the varying local material properties is crucial to accurately model nonlinear and even failure behaviors. For example, even a small temperature (or electric potential) gradient can result in locally very ``hot'' spot (or high electric current spot) due to the gradient concentration effects, leading to material failure in that region \cite{sahimi2003heterogeneous}. The preponderance of previous studies of field fluctuations in heterogeneous material mainly focused on the overall distribution of field values (i.e., the density of states) \cite{cheng1997electric, cule1998electric} and very few studies have investigated the spatial fluctuations and their connection to the fluctuations of the material properties. 

We first present our numerical results of the fluctuating part of the temperature fields resulted from different unitary macroscopic temperature gradients. Subsequently, we present a comprehensive perturbation analysis to rigorously connect the fluctuations of the temperature field to those of the local conductivity field via their respective spectral density functions. We show that in the weak-contrast limit, i.e., when the fluctuations of the conductivity are much smaller than the average value, the temperature fields associated with Class-I hyperuniform materials (characterized by $\alpha \ge 2$) are also hyperuniform, albeit with a lower hyperuniformity exponent ($\alpha-2$). As one moves away from this limit, the fluctuations of the temperature field develop a diverging spectral density at the origin, revealed by our higher-order analysis.

%in heterogeneous materials where the thermal conductivity has hyperuniform perturbations. Hyperuniformity implies an anomalously low variance at large length scales (suppressed long-wavelength fluctuations in the conductivity field). To explore this, we apply different macroscopic temperature gradients and vary the hyperuniform exponent $\alpha$ that characterizes the conductivity field’s spectral decay.

\subsection{Numerical Results}

Figures~\ref{fig:1} to \ref{fig:3} show the steady-state temperature field (upper panels) $T({\bf x})$, temperate fluctuations (middle panels) $T_p({\bf x})$, and the associated spectral density functions \(\tilde{\chi}_{_{T}}(\mathbf{k})\) (lower panels) for various $\alpha$ values, resulted from unitary applied gradients ${\bf G}$ along the horizontal, vertical and diagonal directions, respectively. In these calculations, we have employed a conductivity field $\mathcal{K}({\bf x}) = \mathcal{K}_0 + \delta \mathcal{K}({\bf x})$, where $ \delta \mathcal{K}({\bf x})$ possesses zero mean and unitary variance, and $\mathcal{K}_0 = 5$.

It can be clearly seen that the direction of the imposed macroscopic gradient ${\bf G}$ influences the temperature fluctuation patterns $T_p({\bf x})$ and their spectral density functions \(\tilde{\chi}_{_{T}}(\mathbf{k})\). In particular, although the local conductivity field $\mathcal{K}({\bf x})$ (as shown in Figs. \ref{fig:composite1} and \ref{fig:composite2}) are statistically isotropic, the corresponding temperature fluctuation $T_p({\bf x})$ exhibit clear symmetry breaking patterns correlated with the direction of the applied gradient ${\bf G}$. For example, the clustering regions of positive and negative $T_p({\bf x})$ tend to form elongated patterns perpendicular to the applied gradient. This feature becomes more apparent with decreasing $\alpha$, i.e., as the conductivity field becomes less hyperuniform. This broken symmetry in $T_p({\bf x})$ is also evident in the associated spectral density functions \(\tilde{\chi}_{_{T}}(\mathbf{k})\), which exhibit a minimal two-fold symmetry corresponding to the direction of the applied gradient.

%Comparing Figures~\ref{fig:1}, \ref{fig:2}, and \ref{fig:3}, the temperature perturbation field is not statistically isotropic; instead, structures tend to align with or reflect the orientation of the driving gradient. For example, under a horizontal gradient (Figure~\ref{fig:1}), the perturbation field may exhibit elongated features in the horizontal direction, whereas a vertical gradient (Figure~\ref{fig:2}) produces vertically oriented features. A diagonal gradient (Figure~\ref{fig:3}) leads to a more complex pattern influenced by both $x$ and $y$ directions. This symmetry breaking is also evident in the spectral density, which shows anisotropic distribution of energy in Fourier space depending on the direction of the applied gradient.

In addition, as $\alpha$ decreases, the temperature field develops increasingly pronounced large-scale fluctuations. $\mathcal{K}({\bf x})$ with higher $\alpha$ yields relatively smooth and spatially uniform perturbations in $T({\bf x})$; whereas $\mathcal{K}({\bf x})$ with lower $\alpha$ produces much larger fluctuations in $T({\bf x})$ manifested as more heterogeneous large ``hot'' and ``code'' regions spanning the domain. Analysis of the associated \(\tilde{\chi}_{_{T}}(\mathbf{k})\) reveals a growing peak at the center (zero wavenumber) as $\alpha$ decreases, consistent with the increasing fluctuations in the temperature field. This suggests that the temperature fluctuations can develop significant large-scale fluctuations manifested as a diverging spectral density at the zero-wavenumber limit, even for $\mathcal{K}({\bf x})$ with small $\alpha$. 

%From these numerical observations, several key points are evident:
%\begin{itemize}
%    \item \textbf{Enhanced Large-Scale Fluctuations:}     
%    \item \textbf{Spectral Density Peaks at Low Wavenumber:} 
%\item \textbf{Anisotropy and Symmetry Breaking:} 
%\end{itemize}

\subsection{Perturbation Analysis}

To better understand these results, we carry out a comprehensive perturbation analysis on the heat conduction equation. In particular, we consider the steady-state heat equation with spatially varying thermal conductivity:
\begin{equation}
-\nabla\cdot\Bigl(\mathcal{K}(\mathbf{x})\nabla T(\mathbf{x})\Bigr)=0 \quad \text{in } \Omega,
\end{equation}
with
\begin{equation}
\mathcal{K}(\mathbf{x}) = \mathcal{K}_0 + \delta\mathcal{K}(\mathbf{x}),
\end{equation}
where $\mathcal{K}_0$ is the average conductivity and $\delta\mathcal{K}(\mathbf{x})$ is the fluctuating perturbation with zero mean and unitary variance and spectral density \(\tilde{\chi}_{_\mathcal{K}}(\mathbf{k})\). Without loss of generality, We assume a uniform macroscopic temperature gradient $\mathbf{G}$ is imposed (so that in the absence of heterogeneity, $T_0(\mathbf{x}) = -\,\mathbf{G}\cdot \mathbf{x}$ is the linear background temperature). We seek the solution in an expansion form: 
\begin{equation}
T(\mathbf{x}) = T_0(\mathbf{x}) + \epsilon T_1(\mathbf{x}) + \epsilon^2 T_2(\mathbf{x}) + \mathcal{O}(\epsilon^2),
\end{equation}
where $T_1$ and $T_2$ respectively correspond to the first and second-order perturbations due to $\delta\mathcal{K}(\mathbf{x})$, and $\epsilon \sim \delta \mathcal{K}/\mathcal{K}_0$ is the property contrast parameter. 

In first-order (linear) perturbation theory, i.e., in the weak-contrast limit with very small $\epsilon$, the temperature perturbation $T_1$ satisfies:
\begin{equation}
\nabla \cdot \bigl[\mathcal{K}_0 \,\nabla T_1(\mathbf{x})\bigr] \;+\; \nabla \cdot \bigl[\delta\mathcal{K}(\mathbf{x}) \,\nabla T_0(\mathbf{x})\bigr] \;=\; 0.
\end{equation}
Because $T_0$ has a constant gradient $\nabla T_0 = -\,\mathbf{G}$, this simplifies to a Poisson equation for $T_1$:
\begin{equation}
\mathcal{K}_0 \,\nabla^2 T_1(\mathbf{x}) 
\;=\; \mathbf{G} \cdot \nabla \bigl[\delta\mathcal{K}(\mathbf{x})\bigr].
\end{equation}
Taking the Fourier transform of the above equation yields
\begin{equation}
\mathcal{K}_0\,|\mathbf{k}|^2 \,\widehat{T_1}(\mathbf{k}) 
\;=\; i\,\bigl[\mathbf{k}\cdot \mathbf{G}\bigr]\;\widehat{\delta\mathcal{K}}(\mathbf{k}).
\end{equation}
where we used hats to denote Fourier transforms.
Solving for $\widehat{T_1}(\mathbf{k})$ gives:
\begin{equation}
\widehat{T_1}(\mathbf{k}) 
\;=\; -\,\frac{i\,(\mathbf{k}\cdot \mathbf{G})}{\mathcal{K}_0\,|\mathbf{k}|^2}\;\widehat{\delta\mathcal{K}}(\mathbf{k}).
\end{equation}
This expression shows that the induced temperature fluctuations associated with wavevector $\mathbf{k}$ are proportional to the conductivity fluctuation with the same wavevector, albeit rescaled by a factor 
\(
-\,(\mathbf{k}\cdot \mathbf{G})/ \bigl(\mathcal{K}_0 \,|\mathbf{k}|^2\bigr).
\) 
The factor $(\mathbf{k}\cdot \mathbf{G})/|\mathbf{k}|^2$ encodes two important effects: (i){\it Spectral weighting:} fluctuations with smaller $|\mathbf{k}|$ (large-scale variations) are amplified by the $1/|\mathbf{k}|^2$ term, suggesting that, all else being equal, the temperature field is more sensitive to long-wavelength conductivity fluctuations than to short-wavelength ones. (ii) {\it Anisotropy induced by $\mathbf{G}$:} The dot product $(\mathbf{k}\cdot \mathbf{G})$ indicates that only the component of $\mathcal{K}({\bf k})$ along the gradient ${\bf G}$ contributes to $T_1$, leading to the observed symmetry breaking effects in the temperature flucutations. 

%This implies that fluctuations oriented parallel to the macroscopic gradient have the strongest effect, which breaks isotropy if $\mathbf{G}$ picks a preferred direction.

The spectral density of the first-order temperature field is related to that of the conductivity field
\begin{equation}
\tilde{\chi_{_\mathcal{K}}}(\mathbf{k}) = \bigl|\widehat{\delta\mathcal{K}}(\mathbf{k})\bigr|^2,
\end{equation}
via the following relation:
\begin{equation}
\tilde{\chi}_{_{T_1}}(\mathbf{k}) \;=\; \bigl|\widehat{T_1}(\mathbf{k})\bigr|^2 
\;\approx\; \frac{\bigl(\mathbf{k}\cdot \mathbf{G}\bigr)^2}{\mathcal{K}_0^2\,|\mathbf{k}|^4}\;\tilde{\chi_{_\mathcal{K}}}(\mathbf{k}).
\label{eq_chi_T1}
\end{equation}
For a hyperuniform conductivity field, $\tilde{\chi}_{_\mathcal{K}}(\mathbf{k})$ is anomalously suppressed as $|\mathbf{k}|\to 0$. In our hyperuniform materials, $\tilde{\chi}_{_\mathcal{K}}(k)\sim k^\alpha$ for small $k$, where $\alpha > 0$. According to the first-order theory Eq. (\ref{eq_chi_T1}), for $\alpha > 2$, $\tilde{\chi}_{_{T_1}}(k) \sim k^{\,\alpha - 2}$ will tend to zero as $k \to 0$, indicating that the associated temperature field would also be hyperuniform in this weak contrast limit. On the other hand, if $\alpha < 2$, $\tilde{\chi}_{_{T_1}}(k)$ would diverge in the zero-$k$ limit, indicating large-scale temperature fluctuations persist and the field is not hyperuniform. 

%Thus, first-order theory predicts that conductivity perturbations with $\alpha>2$ should not induce large-scale temperature variance, aligning with the notion of hyperuniform transport behavior in the linear regime.

\begin{figure*}[htbp]
    \centering
    \includegraphics[width=0.6\textwidth]{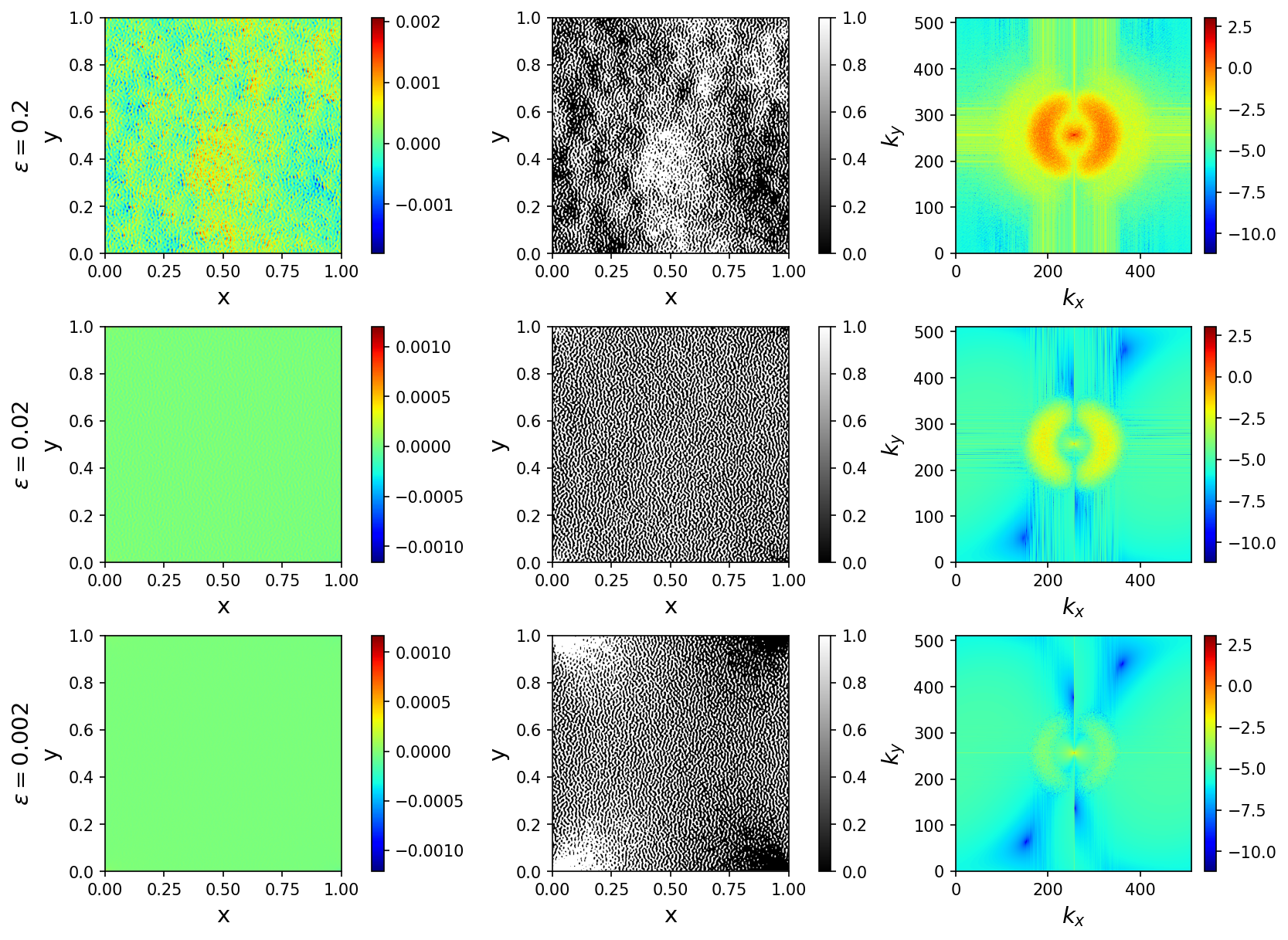}
    \caption{Steady-state temperature fields (left panels), the corresponding binary fields for better visualization (middle panel) and the corresponding log-scale spectral density functions (right panels) for $\alpha = 20$ and $\epsilon = 0.2$, 0.02 and 0.002 (from top to bottom).}
    \label{fig:composite3}
\end{figure*}

% Composite images for \(\alpha=20\) showing the temperature field and its log spectral density for different baseline conductivities: top panel \(\sigma_0=50\), middle panel \(\sigma_0=500\), and bottom panel \(\sigma_0=5000\). Each composite panel includes both the temperature fluctuations and the associated spectral density.

Note that our numerical results with $\mathcal{K}_0$ and unitary variance $\delta \mathcal{K}({\bf x})$ leads to $\epsilon = 0.2$, and the temperature fluctuations even with large $\alpha \ge 10$ show a clear deviation from this first-order prediction and exhibit a significant low-$k$ component in their spectral density function. To further assess the validity of the first-order (linear) perturbation theory, we perform a series of additional numerical simulations with $\alpha = 20$ and $\epsilon = 0.2$, 0.02 and 0.002. Figure~\ref{fig:composite3} shows the steady-state temperature fields (left panels), the corresponding binary fields for better visualization (middle panel) and the corresponding log-scale spectral density functions (right panels) for these three cases. It can be clearly seen that as $\epsilon$ decreases, i.e., the weak contrast limit is approached, the temperature field becomes more uniform and the associated spectral density shows a significantly more suppressed value at the origin. In particular, the temperature field with $\epsilon \sim 0.002$ is already effectively hyperuniform, possessing a very small value of $\tilde{\chi}_{_{T_1}}(k\rightarrow 0)$. These results indicates the validity of our first-order analysis.

%When \(\sigma_0=50\) (panels (a) and (b)), the ratio of the random fluctuation amplitude to the mean is high, pushing the system beyond the strict linear regime. The temperature perturbations exhibit large-scale clustering of hot and cold regions, indicating that second-order (and higher-order) effects become influential and redistribute energy into long-wavelength modes despite \(\alpha=20\).

%Panels (c) and (d) (\(\sigma_0 = 500\)) correspond to a moderate scenario: the same random fluctuations are now a smaller fraction of the mean, and the overall conductivity field is more uniform. Consequently, the induced temperature field exhibits only moderate clustering, and the low-wavenumber power is less pronounced, reflecting partial recovery of hyperuniform behavior.

%Finally, panels (e) and (f) (\(\sigma_0 = 5000\)) show that when the baseline conductivity dominates, the added random fluctuations are negligible. The system approaches near-homogeneity, resulting in very faint temperature perturbations and an almost completely suppressed low-wavenumber power spectrum, in line with the predictions of first-order theory.

%This discrepancy indicates that higher-order effects are important.

To better understand the divergence of $\tilde{\chi}_{_{T_1}}(k\rightarrow 0)$ as one moves away from the weak contrast limit, we proceed with the second order perturbation analysis, focusing on the temperature perturbation influenced by quadratic terms in $\delta\mathcal{K}$. Collecting terms of order $(\delta\mathcal{K})^2$, we obtain
\begin{equation}
\mathcal{K}_0 \,\nabla^2 T_2(\mathbf{x}) \;+\; \nabla \cdot \Bigl[\delta\mathcal{K}(\mathbf{x}) \,\nabla T_1(\mathbf{x})\Bigr] \;=\; 0.
\end{equation}
Taking the Fourier transform, the divergence of the product $\delta\mathcal{K} \,\nabla T_1$ becomes a convolution, so that
\begin{equation}
\mathcal{K}_0\,|\mathbf{k}|^2\,\widehat{T_2}(\mathbf{k}) 
\;=\; i \int d\mathbf{q}\;\Bigl[(\mathbf{k} - \mathbf{q})\cdot \mathbf{G}\Bigr]\,\widehat{\delta\mathcal{K}}(\mathbf{q})\,\widehat{\delta\mathcal{K}}\bigl(\mathbf{k} - \mathbf{q}\bigr),
\end{equation}
where $\mathbf{G}$ denotes the imposed macroscopic gradient. This convolution implies that pairs of conductivity fluctuations at wavevectors $\mathbf{q}$ and $\mathbf{k}-\mathbf{q}$ interact nonlinearly to produce a temperature fluctuation at $\mathbf{k}$. 

Importantly, when one computes the spectral density of the second-order flucutations in the temperature field, $\tilde{\chi}_{_{T_2}}(\mathbf{k}) = \mathbb{E}\bigl[|\widehat{T_2}(\mathbf{k})|^2\bigr]$, the nonlinear interaction leads to an expression that involves the self-convolution of the conductivity spectrum:
\begin{equation}
\tilde{\chi}_{_{T_2}}(\mathbf{k}) \propto \frac{1}{\mathcal{K}_0^2\,|\mathbf{k}|^4} \int d\mathbf{q}\; \Bigl[(\mathbf{k} - \mathbf{q})\cdot \mathbf{G}\Bigr]^2 \,\tilde{\chi}_{_\mathcal{K}}(\mathbf{q})\,\tilde{\chi}_{_\mathcal{K}}(\mathbf{k}-\mathbf{q}),
\label{eq_chi_T2}
\end{equation}
where $\tilde{\chi}_{_\mathcal{K}}(\mathbf{k}) = \bigl|\widehat{\delta\mathcal{K}}(\mathbf{k})\bigr|^2$. Even if $\tilde{\chi}_{_\mathcal{K}}(\mathbf{k})$ is strongly suppressed at low wavenumbers due to hyperuniformity, the convolution can reintroduce significant spectral power at small $|\mathbf{k}|$. The product of two higher-$k$ components can yield a non-negligible contribution near $\mathbf{k}=0$. Along with the $1/|{\bf k}|^4$ factor, this explains the observed low-wavenumber peak in the temperature spectrum. Moreover, the presence of the factor $\bigl[(\mathbf{k} - \mathbf{q})\cdot \mathbf{G}\bigr]$ further enhances the symmetry breaking effects due to applied gradient, leading to a spectral density that is both amplified at small wavenumbers and strongly anisotropic, consistent with the ``butterfly-shaped'' patterns seen in numerical simulations.

%introduces a directional bias: not only are individual modes weighted by their alignment with the macroscopic gradient (as in the first-order term), but the convolution of modes also accentuates anisotropic features in the temperature field. This 

In summary, while the first-order theory predicts hyperuniform temperature fluctuations for $\alpha>2$ in the weak contrast limit, the inclusion of second-order effects reveals that nonlinear mixing transfers spectral power from moderate to long wavelengths (i.e., small wavenumbers). This nonlinear coupling, captured by the convolution integral in Eq. (\ref{eq_chi_T2}), is responsible for the enhanced large-scale fluctuations and anisotropy in the temperature field, explaining the observed simulation results.

\section{Effective Conductivity}
\label{sec:effective_conductivity}

%\subsection{Numerical Homogenization Procedure}

Last but not least, we employ a two-stage numerical approach to compute the effective conductivity. First, we generate realizations of spatially varying conductivity fields with $\mathcal{K}_0 = 5000$ and $\delta \mathcal{K}({\bf x})$ with zero mean and a variance of 10 for different $\alpha$ values (cf. Eq.(\ref{eq:chi})). Second, for each realization of \(\sigma(\mathbf{x})\), we solve the steady-state heat conduction equation (\ref{eq_heat_eq}) under periodic boundary conditions using the finite element method implemented in \texttt{FEniCS}. A small macroscopic temperature gradient is imposed:
\begin{equation}
\mathbf{G} = 
\begin{cases}
(1,\,0) & \text{for computing } \mathcal{K}^e_{xx},\\[1mm]
(0,\,1) & \text{for computing } \mathcal{K}^e_{yy},
\end{cases}
\end{equation}
and the net heat flux is averaged over the domain to obtain the effective conductivity using Eqs. (\ref{eq_Kexx}) and (\ref{eq_Keyy}). We considered a square domain of linear size $L = 50$ with a \(256\times256\) mesh. 8000 independent realizations per \(\alpha\) for each imposed gradient were used and a MUMPS sparse direct solver was employed within \texttt{FEniCS}. 
%Simulations were performed on a high-performance computing cluster via a SLURM array of 400 jobs, each running approximately 20 minutes on a single CPU core, with a total of 133 CPU hours.

%\[
%\sigma_{\mathrm{e},xx} = -\frac{\langle J_x \rangle\,L}{\Delta T}, \quad \sigma_{\mathrm{e},yy} = -\frac{\langle J_y \rangle\,L}{\Delta T},
%\]
%where \(L\) (256 units) is the domain size and \(\Delta T\) is the temperature difference corresponding to the imposed gradient in a homogeneous medium.

%the form
%\[
%\sigma(\mathbf{x}) = \sigma_0 + 10\,\delta\sigma_{\text{rand}}(\mathbf{x}),
%\]
%where \(\delta\sigma_{\text{rand}}(\mathbf{x})\) is a Gaussian random perturbation field with a prescribed power spectrum that decays according to the hyperuniform exponent \(\alpha\). In our study, all simulations are performed using a baseline conductivity of \(\sigma_0 = 5000\).

%\subsection{Results: \(\sigma_{\mathrm{e},xx}\) and \(\sigma_{\mathrm{e},yy}\) vs.\ \(\alpha\)}

\begin{figure}[H]
\centering
\includegraphics[width=0.45\textwidth]{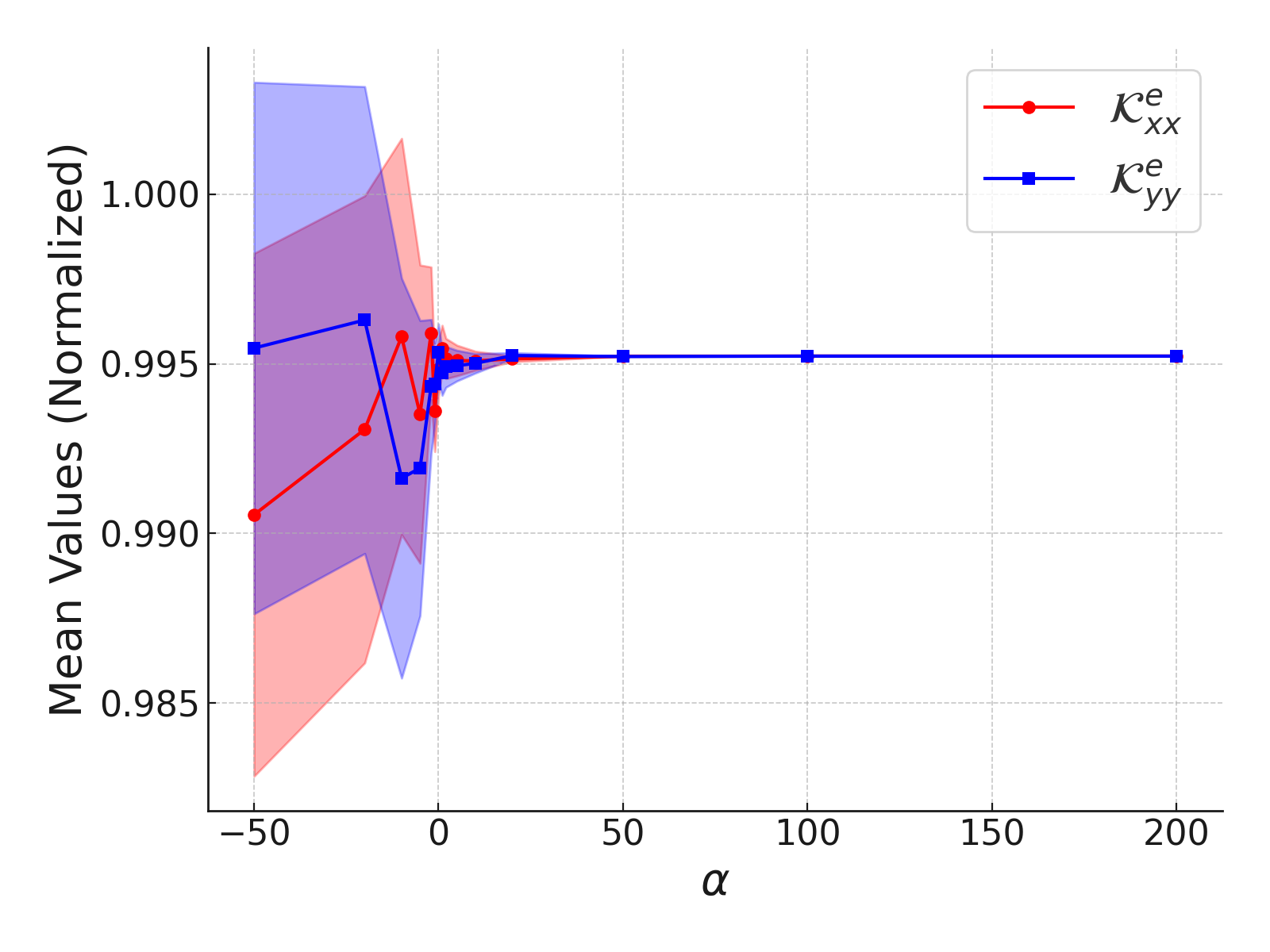}
\caption{Mean values of \(\mathcal{K}^e_{xx}\) (red circles) and \(\mathcal{K}^e_{yy}\) (blue squares) versus \(\alpha\) with 95\% confidence intervals and normalized with respect to $\mathcal{K}_0$.}
\label{fig:mean_eff}
\end{figure}
%For large positive \(\alpha\), both components converge near 4980--5000, indicating nearly homogeneous conduction. For negative or small \(\alpha\), enhanced long-range fluctuations lead to increased variability.

\begin{figure}[H]
\centering
\includegraphics[width=0.45\textwidth]{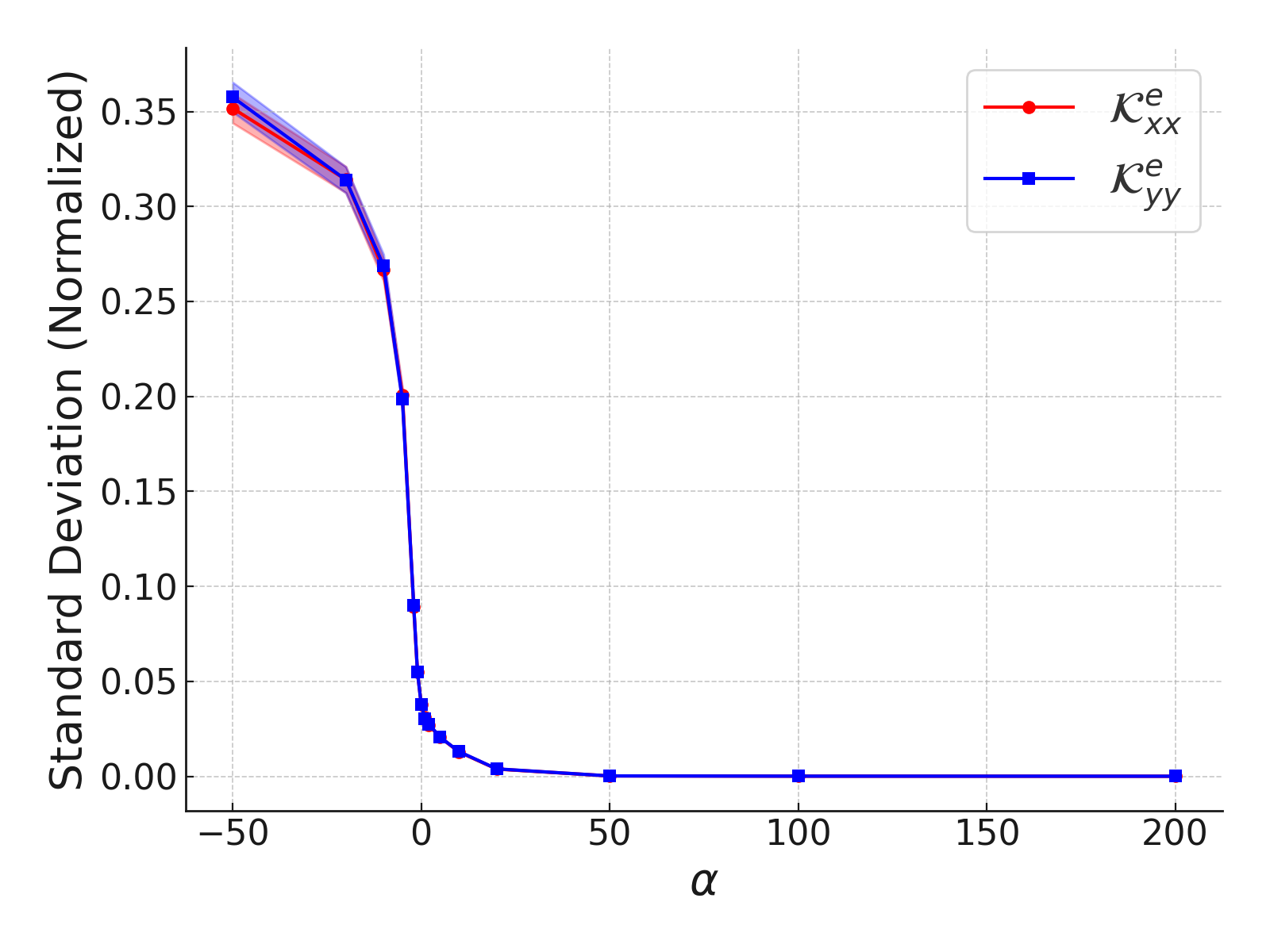}
\caption{Standard deviations of \(\mathcal{K}^e_{xx}\) (red) and \(\mathcal{K}^e_{yy}\) (blue) versus \(\alpha\) with 95\% confidence intervals and normalized with respect to $\mathcal{K}_0$.}
\label{fig:std_eff}
\end{figure}
%. Negative \(\alpha\) results in high variability, while the variability rapidly decreases with increasing \(\alpha\), consistent with the suppression of low-wavenumber fluctuations in hyperuniform media.

%{\color{red} maybe we should normalize the effective conductivity with respect to $K_0$, and also adjust the y coordinate from 0-1.Same applies to the mean}

We investigate fifteen distinct values of \(\alpha\) spanning antihyperuniform to nonhyperuniform to hyperuniform systems, including $\alpha = -50$, -20, -10, -5, -2, -1, 0, 1, 2, 5, 10, 20, 50, 100 and 200. The computed effective conductivity values are statistically analyzed to determine the mean and standard deviations with 95\% confidence intervals of \(\mathcal{K}^e_{xx}\) and \(\mathcal{K}^e_{yy}\) as functions of \(\alpha\), respectively shown in Fig. \ref{fig:mean_eff} and Fig. \ref{fig:std_eff}. It can be seen for hyperuniform systems ($\alpha > 0$), the materials exhibit a high degree of isotropy in terms of effective conductivity with virtually identical values of \(\mathcal{K}^e_{xx}\) and \(\mathcal{K}^e_{yy}\). Increasing $\alpha$ also leads to an increase of the effective conductivity, which saturates beyond $\alpha \ge 20$. On the other hand, the nonhyperuniform/antihyperuniform systems exhibit large anistropy and significant fluctuations of $\mathcal{K}^e$, resulted from large-scale fluctuations in their local conductivity fields.  

It is interesting to note that the standard deviations of the effective conductivity exhibit a sharp transition from high values to vanishing small values as one move from antihyperuniform/nonhyperuniform systems to hyperuniform systems, mimicking a phase-transition-like behavior. The vanishingly small standard deviations of the hyperuniform materials with large $\alpha$ indicate ultra uniformity in the effective properties across realizations of these material systems, which is highly desirable for applications under extreme conditions. On the other hand, antihyperuniform systems can exhibit significantly larger variance across realizations, resulted from their structural fluctuations.

%This is because the microstructures associated with such large $\alpha$  

%\subsection{Physical Insights and Computational Details}

%The trends observed in Figures~\ref{fig:mean_eff} and \ref{fig:std_eff} indicate that:
%\begin{itemize}
%    \item For large positive \(\alpha\), the conductivity field exhibits strong hyperuniformity, and the effective conductivity stabilizes with minimal variability. This uniformity suggests that low-wavenumber fluctuations are effectively suppressed.
%    \item For negative or small \(\alpha\), strong long-range fluctuations in \(\sigma(\mathbf{x})\) introduce significant variability in the effective conductivity, with the formation of extensive regions of higher or lower conductivity that modulate the net heat transport.
%\end{itemize}

%These results demonstrate that enhanced hyperuniformity (large positive \(\alpha\)) leads to robust and reproducible effective conduction, while the lack of hyperuniformity (negative or small \(\alpha\)) results in significant variability in macroscopic transport properties.

%\textcolor{red}{Here we present the structure-property relationship, i.e., $\sigma_e$ vs $\alpha$. First describe the procedures for numerical homogenization with necessary details. Then present $\sigma_e$ vs $\alpha$ plot, and variance vs. $\alpha$ plot. Describe the trend and provide physical insights and explanations. Also need to report the computational details, e.g., number of realizations, system size, physical parameters, CPU hours etc.}

\section{Conclusions and Discussion}

%\textcolor{red}{summarize current study, i.e., investigation of structure-property relationship; distinct from previous work on binary, we model continous field. and the key findings. mention future work, generlaization to other properties, and development of new theory, connection to binary systems.}

In this work, we presented a comprehensive investigation of the structure-property relationship in a class of disordered hyperuniform heterogeneous materials characterized by an analytical spectral density function with power-law small-$k$ scaling. Distinct from the preponderance of previous studies on DHU heterogeneous materials, which focused on two-phase material systems modeled by a binary field, we considered systems possessing continuously varying local material properties $\mathcal{K}({\bf x})$ (e.g., thermal or electrical conductivity), modeled by a random scalar field. We presented a highly effective Fourier filtering generative method to render realizations of the systems and showed that by controlling the scaling exponent $\alpha$, a wide spectrum of distinct material microstructures spanning from hyperuniform ($\alpha>0$) to nonhyperuniform ($\alpha=0$) to antihyperuniform ($\alpha<0$) systems can be obtained. Moreover, we carried a detailed analysis of the physical field fluctuations both numerically and analytically via perturbation theory. We showed that in the weak-contrast limit, i.e., when the fluctuations of the property are much smaller than the average value, the physical fields associated with Class-I hyperuniform materials (characterized by $\alpha \ge 2$) are also hyperuniform, albeit with a lower hyperuniformity exponent ($\alpha-2$). As one moves away from this weak-contrast limit, the fluctuations of the physical field develop a diverging spectral density at the origin and may lose hyperuniformity. We also computed the effective properties of the material systems and establish an end-to-end mapping connecting the scaling exponent $\alpha$ to the overall effective conductivity of the material system via numerical homogenization. We observe a sharp decrease of the variance of effective properties across realizations as $\alpha$ increases from antihyperuniform values to hyperuniform values. These results have significant implications for the design of novel DHU materials with targeted physical properties. 

%\textcolor{red}{make connections to composite design and optimization; manufacturing via 3D printing}

%in ${\tilde \chi}_{_V}({\bf k})$

%Fourier-space based numerical construction procedure in 3D. 

Although focusing on diffusive transport properties, our approach can be immediately generalized to study other physical properties of interest, including elasticity, fluid permeability, and wave propagation properties. In particular, generalizing the perturbation analysis to wave equations would enable one to directly connect the density of states to the spectral density of the material system. It is also of great interest to develop rigorous perturbation expansions that quantitatively connect the effective properties to the hierarchy of statistical descriptors of the material (such as the standard $n$-point correlation functions) characterizing the continuously varying local material properties, which are crucial for inverse material design. In additional, all of our analysis and simulations can be readily generalized to three-dimensional material systems. We will explore these generalizations in our future work.

\noindent{\bf Data Availability Statement}: The codes and data are available upon request.

%at: \textsc{https://github.com/orgs/cmgrouplab/repositories}.

%\begin{acknowledgments}
%This work was supported by the Army Research Office under Cooperative Agreement Number W911NF-22-2-0103.
%\end{acknowledgments}

\smallskip

\bibliography{reference}% Produces the bibliography via BibTeX.

\end{document}